    \documentclass[a4paper,oneside,Palatino]{article}
    \usepackage{latexsym}
    \usepackage[dvips]{color}
    \usepackage[dvips]{graphicx}
    \usepackage[dvips]{graphics}
 \usepackage{amsmath}
    \usepackage{amssymb}
    \usepackage{latexsym}
    \usepackage{cite}
    \usepackage[dvips]{color}
    \usepackage[dvips]{graphicx}

\def\p{\partial}

\def\tsi{\tilde{\psi}} 
\def\si{\psi}

\catcode`@=11

\@addtoreset{equation}{section}
\catcode`@=11

\@addtoreset{equation}{section}
\catcode`@=11

\title{{\Large Signatures Of Scalar Photon Interaction In Astrophysical Situations.}}

\author{\bf  Avijit K. Ganguly\footnote{ Corresponding author}, Manoj K. Jaiswal, Shikha Awasthi.
\thanks{e-mail addresses: avijitk@hotmail.com, manojau.87@gmail.com,
awasthi.shi@gmail.com }\\
\normalsize
Dept. Of Physics. MMV, Banaras Hindu University, Varanasi 221005, India.
\normalsize\\
}

\begin{document}

\maketitle

\begin{center}
\begin{abstract}
\noindent
Dimension-5 photon ($\gamma$) scalar ($\phi$) interaction term usually 
appear in the Lagrangians of bosonic sector of unified theories of 
electromagnetism and gravity. This interaction makes the medium dichoric and 
induces optical activity. 
Considering a toy model of an ultra-cold  magnetized 
compact star ( White Dwarf (WD) or Neutron Star (NS)), we have modeled 
the propagation of very low energy 
photons with such interaction, in the environment of these stars.
Assuming synchro-curvature process as the dominant mechanism of emission in such environments, we have tried to understand the polarimetric implications of 
photon-scalar coupling on the produced spectrum of the same. 
Further more assuming the 'emission-energy vs 
emission-altitude' relation, that is believed to hold in such 
(i.e. cold magnetized WD or NS) environments, we have tried to point out 
the possible modifications to the radiation spectrum when the same is 
incorporated along with dim-5 photon scalar mixing operator. 
\end{abstract}
\end{center} 
\noindent



\renewcommand{\thefootnote}{(\arabic{footnote})}

\section{Introduction.}
\label{sec:intro.}
 Scalar $\phi(x)$ photon $\gamma$ interaction through dimension five operators
originates in many theories beyond standard model of particle physics, 
usually in the unified theories of electromagnetism and gravity \cite{duff}.
The scalars involved can be moduli fields of string theory, KK particles
from extra dimension, scalar component of the gravitational multiplet in 
extended supergravity models etc. to name a few [\!\! \cite{Schmutzer }-\cite{Fischbach}].Though the main emphasis of such models had been unification of forces, however the same idea has also been used to solve the dark matter  and dark energy [\cite{kom}--\!\!\!\cite{dunk}] problems of the universe, as in the 
chameleon models for dark matter[\cite{brax}--\!\!\!\cite{davis}].\\ 

\noindent
Physics of compactification--of extradimensions--introduces various kinds
of model dependent interactions. For instance in  the original Kaluza- Klein 
(KK) model \cite{kaluza}, compactification of the extra fifth dimension was 
constructed to unify gravity with electromagnetism, that introduced  dim-5
$\phi \gamma $ interaction. In string theories--constructed in ten or twenty 
six dimensions--the extra-dimensions are compactified and  models have been 
made to find out various observable consequences of the same. One such model, 
invoked to explain the weakness of gravity relative to other forces, was 
developed by Arkani-Hamed, Dimopoulos and Dvali \cite{ADD}. In 
this model, ( also known as ``Large Extra Dimension'' model), the standard model particles are confined 
to a four dimensional membrane, and gravity propagates in the other spatial dimensions, those are 
large compared to plank scale. However there also exist models ( with one or more extra 
spatial-dimensions ), where all fields propagate universally \cite{ued}. Collider  signatures of Higher 
dimensions has been studied in \cite{exsighd}. Other than the collider signatures mentioned above, 
effort has also been made in \cite{seki}, to relate the effect of extra dimension with four dimensional 
cosmological constant. \\ 

\noindent
Interestingly enough, the dimension-5 photon scalar interaction term 
generated in the Lagrangian \cite{kaluza} by compactification of
the extradimension, would  eventually bear optical signatures of the same. 
The purpose of this note is to study such effects coming from dim-5 scalar 
photon interaction in some detail and shed some new light on their 
astrophysical consequences. \\
\indent
Usually dim-5 interactions (be it  scalar photon or pseudoscalar 
(axion) photon interactions), induce optical activity. The scalar or 
pseudoscalar photon interaction turns vacuum into a birefringent and  
dichoric one \cite{raffelt, Ganguly3}. As a result, the plane of polarization of a plane polarized beam of 
light keeps on rotating as it passes through vacuum. 
This particular aspect of the 
theory has been exploited extensively in the literature in a different physical 
context (with pseudoscalar photon coupling) [\cite{jain1}-\cite{criticism}].\\
 
\noindent
Keeping this in view, in this note, initially, we have studied the 
optical signatures of  dim-5, $\phi F_{\mu\nu}F^{\mu\nu}$ interaction 
in an ambient magnetic field of strength $ \sim 10^{13}$ Gauss. 
Following this, we have considered a toy model for the steller enviornment
of a strongly magnetized, rotating, ultra cold, compact astrophysical objects
 like White Dwarf (WD) and Neutron Star (NS)--to understand some  
issues related to--emission energy and altitude \cite{ee1}, that is believed
to affect the produced spectra of electromagnetic radiations (EM), 
originating  there. As a next logical step, we have made an effort to explore 
the potential of the later, in modifying the usual spectral signatures of the 
tree-level dim-5, $\phi F_{\mu\nu} F^{\mu\nu}$ coupling--from such astrophysical 
situations.  We would like to mention here, that, this model of ours is a toy 
model.  The purpose of this construction is, to motivate further investigations
of the possible modifications to the emission spectra, from actual WD or NS 
environment, when  $\phi F_{\mu\nu} F^{\mu\nu}$ interaction has been taken
into account.\\ 

\noindent 
 While investigating the effects of $\phi F_{\mu\nu}F^{\mu\nu}$ 
interaction,  in this note we have been able to achieve three objectives: 
(i) verifying the earlier results \cite{Ganguly3} through an independent 
approach, (ii) Showing the  possibility of generation of circular and 
elliptic polarization from a plane polarized light beam--as the same
passes through the steller environment.(iii) Pointing out the possibility 
of some extra modifications to the usual polarimetric signatures of 
$\phi F_{\mu\nu}F^{\mu\nu}$ coupling, due to emission altitude vs energy 
relation, that has been discussed in the literature \cite{ee1}.\\ 

\noindent
It became very obvious for the first time in \cite{Ganguly4} that, existence
of superluminal propagation modes for low frequency photons are possible, in 
a model with dimension-5 $\phi F_{\mu\nu}F^{\mu\nu}$ interaction term present 
in the Lagrangian. The analysis there, was performed in terms of the gauge potentials, using Lorentz gauge, leaving a scope of {\it{gauge ambiguities}} as a 
source of the  problem. In order to rule out any role of the same, for the 
afore mentioned problem,  we have taken a different approach, using the field 
strength tensors and Bianchi identity for deriving the same. Our new approach 
establishes the results obtained earlier.\\

\noindent
To shed light on the second issue, we have assumed the radiation, 
produced at the production point, to be plane polarized electromagnetic wave 
( with orthogonal planes of polarization, as is the case for synchro-curvature 
radiation). What  we find is: {\it{the same generates a significant amount 
of scalar component through $\phi F_{\mu\nu}F^{\mu\nu}$ interaction, well before 
it is out of the stellar atmosphere}} ; this is some thing, that needs to be 
considered when one is looking for signatures of dim-5 mixing operators from 
astrophysical polarimetric data. Also, though the initial beam of 
radiation is plane polarized, however during its passage it picks up 
significant amount of elliptic/circular polarization through  mixing. 
The amount of elliptic/circular polarization generated through mixing is 
energy $\omega$ dependent with a complex dependence on $\omega$ 
along with the strength of the magnetic field ${\cal{B}}$, coupling constant
$g_{\gamma\gamma \phi}$ as well as the distance $z$ traveled in the stellar 
atmosphere. During our analysis, we have also looked into the pattern of 
polarization angle $\Psi$ and ellipticity angle $\chi$  that the beam of 
radiation generates at different wavelengths, after  propagation  through 
the same distance. What is very interesting is the  existence of identical 
 polarization angle ($\Psi$) at and ellipticity angle $\chi$ for multiple 
values of energies($\omega$) when the traversing path is same. The details 
of the same are discussed later.\\

\noindent
The organization of the document is as follows, in section-II, we have derived 
the equations of motions. Section three is dedicated to the determination of 
the dispersion relations and the solutions of the equations of motions. In 
section four we discuss about the possible observables and applications, 
including a brief introduction to stokes parameters. Introduction to the 
physics of magnetized astrophysical compact objects ( toy model ) and ideas 
behind energy vs emission altitude mapping [\!\!\cite{ee1}-\cite{ee4}] is 
presented in section five. Results are presented in section VI. Lastly we 
conclude by pointing out the relevance of our analysis in realistic 
astrophysical or cosmological contexts.\\ 
    
\section{From The Action To The Equations Of Motion.}

\noindent
To bring out the essential features of $\phi F_{\mu\nu}F^{\mu\nu}$ coupling 
term on the dynamics of the system,  we would work in flat four dimensional 
space time. The action for this coupled scalar photon system 
in flat four dimensional space time  is given by:\\
 
\begin{eqnarray}\nonumber
S \!=\!\! \int d^4x  \!\left[\frac{1}{2}(\partial{_\mu}) \phi (\partial^{\mu}\phi) 
-\frac{1}{4}g_{\phi\gamma\gamma} \phi F_{\mu\nu}F^{\mu\nu}  
-\frac{1}{4}F_{\mu\nu}F^{\mu\nu} 
 \!\right] \nonumber \\
\label{action1}
\end{eqnarray}
\noindent
 Here $g_{\gamma\gamma\phi}$ is the effective coupling strength between scalar 
and electromagnetic field. The equations of motion can be obtained by varying 
the action with respect to $\phi$ and $F_{\mu\nu}$ and demanding invariance of
the same under this arbitrary variation; the result is:,

\begin{eqnarray}
 \partial_{\mu}\left[F^{\mu\nu} 
+ g_{\phi\gamma\gamma}\phi F^{\mu\nu}
\right]=0 \nonumber \\
\partial_{\mu}\partial^{\mu} \phi+\frac{1}{4}g_{\phi\gamma\gamma}
F_{\alpha\beta}F^{\alpha\beta}=0
\label{eom1}
\end{eqnarray}
As a next step we decompose the EM field into two parts, the mean field 
 (${\bar{F}_{\mu\nu}}$) and  the infinitesimal fluctuation ($f_{\mu\nu}$ ), i.e. :
\begin{eqnarray}
F_{\mu\nu}={\bar{F}_{\mu\nu}}+f_{\mu\nu}.
\end{eqnarray}
Assuming the magnitude of the
scalar field to be of the order of the fluctuating electromagnetic 
field $f_{\mu\nu}$ one can linearize the eqns in [\ref{eom1}] 
\footnote{We further assume that only the $12$ component of the mean field is  
nonzero and rest are zero. Therefore 
\begin{equation}
\bar{F}_{\mu\nu}\tilde{\bar{F}}^{\mu\nu}=0.
\label{EB0}
\end{equation}
}.
\noindent
The equations of motions for the
scalar and the electromagnetic fields turn out to be,\\
%
\begin{eqnarray}
 \partial_{\mu}f^{\mu\nu} 
=-g_{\phi\gamma\gamma} \partial_{\mu}\phi\bar{F}^{\mu\nu}. 
\label{emeq1}
 \\ 
\partial_{\mu}\partial^{\mu} \phi  =- \frac{1}{2} g_{\phi\gamma\gamma} \bar{F}^{\mu\nu}f_{\mu\nu}
\label{emeq2}
\end{eqnarray}
\noindent
The two equations [\ref{emeq1}] and [\ref{emeq2}] are the two equations of 
motion governing the dynamics of the system in an external magnetic field described by the tensor $\bar{F}^{\mu\nu}$. Here we consider $\bar{F}^{\mu\nu}$ to be a
very slowly varying function of coordinates, so that the same can be considered to be effectively constant. We note here that these two equations carry the information about the three degrees of freedom of the system, two for the two 
polarization states of the photon and the third one for the scalar degree 
of freedom.
\section{Dispersion Relation.}
\noindent
In this section we would get the dispersion relation for the scalar photon
coupled system of equations. Equation [\ref{emeq1}] in general would provide
three equations, corresponding to two transverse and one longitudinal states 
of  polarization  for the photon. However in vacuum photons has only two 
transverse degrees of freedom, so one of the three would be redundant. 
The last equation i.e.,[\ref{emeq2}] would provide the dynamics of scalar
degree of freedom, thus making the total number of degrees of freedom 
for the coupled system to be  three.\\ 

\noindent 
We can get to the dynamics of the three degrees of freedom for the scalar 
photon system by two methods, by (a)  using the gauge potentials and choosing
a particular gauge, or by (b) making use of the Bianchi identity. In this work
we would chose the second method. That is using the Bianchi identity:
\\
\begin{eqnarray}
\partial{_\mu}f_{\nu\lambda}+\partial{_\nu}f_{\lambda\mu}+\partial{_\lambda}f_{\mu\nu}=0
\label{bianch}.
\end{eqnarray}
%
\noindent
If we now multiply the Bianchi Identity by $\bar{F}^{\nu\lambda}$ and operate with $\partial^{\mu}$, we arrive at the identity,
%
\begin{eqnarray}
\partial_{\mu}\partial^{\mu} (f_{\lambda\rho} \bar{F}^{\lambda\rho})=-2\partial^{\lambda}\partial_{\mu}(f^{\mu\rho}
\bar{F}_{\lambda\rho}).
\label{bianchi1}
\end{eqnarray}

\noindent
One can now take equation, eqn.(\ref{emeq1}) and multiply it by $\bar{F}_{\nu\lambda}$ and operate with 
$\partial^{\lambda}$ subsequently use eqn. (\ref{bianchi1}), to get to,
\def\pk{k}
\begin{eqnarray}
\p_{\mu} \p^{\mu}(\frac{f\bar{F}}{2}) = g_{\phi\gamma\gamma}
\p^{\lambda}\p_{\alpha}\phi (\bar{F}^{\alpha\nu}\bar{F}_{\nu\lambda})
\label{eom1a}
\end{eqnarray} 

\noindent
Next we introduce a new variable, $\psi=f^{\nu\lambda}\bar{F}_{\nu\lambda}$, use it in  eqn. (\ref{eom1a}) and 
go to momentum space. The resulting equation in momentum space is,\\
%
\begin{eqnarray}
 k^2\psi = g_{\phi\gamma\gamma}
 \left(\pk_{\alpha}\bar{F}^{\alpha\nu}\bar{F}_{\nu\lambda}\pk^{\lambda}\right)\phi
\label{eom1af}
\end{eqnarray}\\

\noindent
Similarly, defining, $\tilde{\psi}= f_{\mu\nu}\tilde{F}^{\mu\nu}$  and using same procedure one arrives at the equation for $\tilde{\psi}$. The same turns out to be,\\
%
\begin{eqnarray}
 k^2 \tilde{\psi} =0
\label{eom2}
\end{eqnarray}

\noindent
Finally the equation of motion for the scalar field in momentum space is 
given by,\\
\begin{eqnarray}
k^2\phi = g_{\phi\gamma\gamma}\psi .
\label{eomphi}
\end{eqnarray}

\noindent
Assuming, $\bar{F}^{12} \ne 0$ therefore $\tilde{\bar{F}}^{03}\ne 0$ and denoting $\bar{F}^{12}=\cal{B}$,
we may use the following compact representations for various Lorentz scalars 
e.g.,
\def\cb{\cal{B}}
$\left(\bar{F}^{\mu\nu}\bar{F}_{\mu\nu} \right) = 2{\cb}^2,$ 
%
 $\left(\pk_{\alpha}\bar{F}^{\alpha\nu}\bar{F}_{\nu\lambda}\pk^{\lambda}\right)= k^2_{\perp}{\cb}^2 $ and
%
$\left(k_{\alpha}\tilde{\bar{F}}^{\alpha\nu}\tilde{\bar{F}}_{\nu\lambda}k^{\lambda}\right)=\left(k^2+k^2_{\perp}\right){\cb}^2 $
appearing in the equations of motion. In these expressions $k_{\perp}$ is the 
component of $\vec{K}$ that is orthogonal to ${\cal{B}}$ and $\Theta$ is the 
angle between the magnetic field {\cal{B}} and the propagation direction 
$\vec{K}$. In terms of these we can rewrite the following expressions as:\\

\begin{eqnarray}
  k^2_{\perp}{\cb}^2 = K^2 Sin^2 (\Theta) {\cb}^2 &\simeq& \omega^2 Sin^2 (\Theta) {\cb}^2 \nonumber \\
 \left(k^2+k^2_{\perp}\right){\cb}^2 = \left( \omega^2- K^2 Cos^2(\Theta) \right) {\cb}^2 
&\simeq& \left( \omega^2 Sin^2(\Theta) \right) {\cb}^2.
\label{kpetc}
\end{eqnarray}

\noindent
While deriving the expressions in eqn. [\ref{kpetc}], we have assumed that, 
to order $g_{\gamma\gamma\phi}$, $\omega \simeq K$.\\

\noindent
Being armed with these (i.e. findings of eqn. [\ref{kpetc}]), the equations 
of motions for the combined photon and scalar system can be written 
in matrix form:\\
%
\begin{eqnarray} 
   \left(\begin{matrix}
     & k^2   \,\,\,   &0  \,\,\, & 0  \cr
     & 0      \,\,\,    & k^2  \,\,\,  &-g_{\phi\gamma\gamma}\omega^2 
                                                                      Sin^2 \Theta {\cb}^2  \cr
&0 \,\,\,   & -g_{\phi\gamma\gamma}  \,\,\,  & k^2
\end{matrix}
\right)
\left(
\begin{matrix}
{\tilde{\psi}} \cr
\psi \cr
\phi
\end{matrix}
\right)= 0
\label{m1}
\end{eqnarray}
%
\noindent
The matrix equation  [\ref{m1}] does not look symmetric because 
the dimension of $\phi$ and the dimensions of $\psi$ or $\tilde{\psi}$ 
are different. To bring the same in symmetric form, we multiply 
the  $\phi$  equation 
in eqn.[\ref{eomphi}] 
by $\omega Sin\Theta {\cal{B}}$, and redefine $\phi$ by $\Phi$, when 
$\Phi =\omega Sin\Theta {\cal{B}}\phi$ to
arrive at, 
%
\begin{eqnarray} 
\left[ 
  \begin{matrix}
     & k^2          & 0                                                                    & 0                                                          \cr
     & 0          & k^2                                                                    &-g_{\phi\gamma\gamma}\left(\omega Sin\Theta {\cal{B}}\right)     \cr
     & 0          &-g_{\phi\gamma\gamma}\left(\omega Sin\Theta {\cal{B}} \right)                & k^2 
  \end{matrix}
\right]
\left[
\begin{matrix}
{\tilde{\psi}} \cr
\psi \cr  
\Phi
\end{matrix}
\right]= 0
\label{m2}
\end{eqnarray}\\

\subsection{Inhomogeneous Wave Equation.}
It can be seen form eqn.[\ref{m2}], that because of the presence of off 
diagonal elements, two dynamical degrees of freedom out of the three,
( $\psi,~ \tilde{\psi}$ and $\Phi$) during their propagation mix with each other. 
The matrix in eqn.[\ref{m2}] is real symmetric so we can go to a diagonal basis, 
by an orthogonal transformation to diagonalize the same. The orthogonal transformation
matrix is given by,
\begin{eqnarray}
{\bf{O}} = \left( \begin{matrix}
                 & 1  \,\,\,   &0  \,\,\, & 0  \cr
                 & 0  \,\,\,   & cos \theta  \,\,\,  &  sin \theta  \cr
                 &0 \,\,\,   & -sin \theta  \,\,\,  &  cos \theta 
                   \end{matrix}
            \right)\!, 
\rm{~where}~ \theta= \frac{\pi}{4}, \rm {~for~the~case~in~hand}.
\end{eqnarray}
\noindent
On diagonalizing eqn. (\ref{m2}), we arrive at,
\begin{eqnarray}
\left(\begin{matrix}
     & k^2  \,\,\,   &0  \,\,\, & 0  \cr
&0 \,\,\,    & k^2-g_{\gamma\gamma\phi}{\cal{B}}sin\Theta\omega  \,\,\,  &  0 \cr
&0 \,\,\,   &  0 \,\,\,  &  k^2+g_{\gamma\gamma\phi}{\cal{B}}sin\Theta\omega
\end{matrix}
\right)
\left(
\begin{matrix}
\tilde{\psi} \cr
\frac{ \Phi+\psi }{\sqrt{2}} \cr
\frac{\Phi -\psi}{\sqrt{2}}
 \end{matrix}
\right)=0
\label{diagbasis1}
\end{eqnarray}
It is easy to see from the equation above that,
$\tilde{\psi}$, $\frac{\psi + \Phi}{\sqrt{2}}$ and
$\frac{-\psi + \Phi}{\sqrt{2}} $ satisfies the following dispersion 
relations,
\begin{eqnarray} 
\omega &=& K  
\label{livs}  \\ 
\omega_{+} &=& \pm \sqrt{\left[K^2+g_{\gamma\gamma\phi}{\cal{B}}sin\Theta\omega 
\right]}  
\label{livas} \\
\omega_{-} &=& \pm \sqrt{\left[K^2-g_{\gamma\gamma\phi}{\cal{B}}sin\Theta\omega 
\right]}
\label{livsn}
\end{eqnarray}

\noindent
Where, the quantity $ g_{\gamma\gamma\phi}{\cal{B}}sin\Theta\omega $  depends on the strength of the
external electromagnetic field, scalar photon coupling constant 
$g_{\phi\gamma\gamma}$, $\omega$ and the sine of the  angle $\Theta$ 
between the direction of propagation $~\vec{K}~$ and the magnetic field 
${\cal{B}}$. \\

\noindent
It should be noted that equation [\ref{m2}] or [\ref{diagbasis1}], incorporates all the dispersive features of 
a photon propagating in a magnetized vacuum with dimension-five scalar photon interaction.
One can verify that the dispersion relations obtained from [\ref{m2}] or [\ref{diagbasis1}] are identical 
to those  in \cite{Ganguly3}, provided appropriate limits are taken. \\

\noindent
Its worth noting that eqn. [\ref{m2}] or [\ref{diagbasis1}] actually shows that in the case of scalar photon interaction, 
photons with polarization state perpendicular to the magnetic field ${\cal{B}}$ remains unaffected 
and propagates with speed of light  and the same with polarization state parallel to the magnetic field
couples to the scalar and undergoes modulation. As we will see later that there exists a critical 
energy ($\omega_c$) below of which the perpendicular modes have imaginary $\omega$, hence they would 
be non-propagating. However the perpendicular mode doesn't suffer from this pathological problem, 
hence (for energy below $\omega_c$ ) they would propagate freely. Therefore light coming from distant 
sources with $\omega < \omega_c$ would appear to be linearly polarized provided this type of interaction 
does exist in nature.\\

\noindent
We emphasize here, that, much of our analysis performed in this paper  would have remained 
the same even if we had dim-5 {\cal{pseudoscalar}} {\cal{photon}} interaction, thus making 
it difficult to identify the type of interaction responsible for polarimetric observation 
that is being invoked in this note.\\

\noindent
However, the way out is to note that the parallel and perpendicularly 
polarized components of the photon in these two different kind of interactions ({\it{scalar or pseudoscalar}}) 
interchange  their role in presence of an external magnetic field. 
Hence the polarization state of the linearly polarized light for scalar photon interaction 
would be orthogonal to the same with Axion photon system. Therefore in principle one can 
look for this signature in polarimetric observations to point out the kind of interaction
responsible for the type of signal.\\ 

\subsection{Inhomogeneous Wave Equations.}
\label{soluns}
\noindent
The solutions for the dynamical degrees of freedom in coordinate space can 
be written as,
\begin{eqnarray}
\left(
\begin{matrix}
\tilde{\psi} \cr
cos \theta ~\psi + sin \theta ~\Phi \cr
-sin \theta ~\psi + cos\theta ~\Phi
 \end{matrix}
\right)
=
\left(
\begin{matrix}
A_{0} e^{i\left( \omega t - k.x \right)} \cr
A_{1} e^{i\left( \omega_{+} t - k.x \right)} \cr
A_{2} e^{i\left( \omega_{-} t - k.x \right)} 
\end{matrix}
\right)
\label{solns}
\end{eqnarray}
\noindent
The constants, $A_{0},~A_{1}$~ and $A_{2}$ has to be defined from the boundary 
conditions one imposes on the dynamical degrees of freedom. from 
eqn.[\ref{solns}] the solutions for the dynamical variables
turn out to be,

\begin{eqnarray}
\tilde{\psi} (t, x) &=& A_{0}\,\, e^{i\left( \omega t - k.x \right)} \nonumber \nonumber \\
\psi (t,x) &=& A _{1}\,\,cos \theta \,\, e^{i\left( \omega_{+} t - k.x \right)} - 
 A _{2}\,\, sin \theta \,\,  e^{i\left( \omega_{-} t - k.x \right)} \\ \nonumber
\Phi (t,x) &=& A _{1} \,\, sin \theta \,\,  e^{i\left( \omega_{+} t - k.x \right)} + 
 A _{2} \,\, cos \theta\,\,  e^{i\left( \omega_{-} t - k.x \right)}.
\label{solnforfields}
\end{eqnarray}

\noindent
In the following we consider the following boundary conditions, $\Phi (0,0) = 0$
and $ \psi(0,0) =1 $. With this boundary condition  we have, 
$ \frac{A_{2}}{sin \theta}= -1 $. And angle $\theta=\frac{\pi}{4}$ as has already been stated before. With these conditions the soln for $\psi$ turns out to be,
\begin{eqnarray}
\psi(t, x)= 
\left[ cos^2 \theta \,\, e^{i\left( \omega_{+} t - k.x \right)} + 
  sin^2 \theta \,\,  e^{i\left( \omega_{-} t - k.x \right)} \right].
\label{solnpsi}
\end{eqnarray}

\noindent
Defining, $ a^2_{x} (t)=\left({\rm{\cal{R}}}e \left[\psi (t,0)\right]\right)^2 
+ \left(\rm{\cal{I}}m \left[\psi (t,0) \right] \right)^2$, we get the 
following the form for $\psi \left(t,x \right)$,
%
\begin{eqnarray}
\psi(t,x) = a_{x} (t) e^{i \left( tan^{-1} \left[\frac{ cos^2 \theta \,\, sin \omega_{+} t+ sin^2 \theta \,\, sin \omega_{-} t }{  cos^2 \theta \,\, cos \omega_{+} t+ sin^2 \theta \,\, 
cos \omega_{-} t} \right]- k x\right)}
\label{inhomosoln}
\end{eqnarray}
%
\noindent
A wave equation of this type is usually called, inhomogeneous wave equation. 
The phase velocity for such system, where the solution is represented by, 
$a(t)e^{i\varphi(t) - kx }$ is defined by,
\begin{eqnarray}
\mbox{{\cal{v}}}_p = \frac{1}{K}\frac{\partial \varphi(t)}{\partial t}
\label{vphase}
\end{eqnarray}\\
\indent
For the case under consideration, since $\theta= \frac{\pi}{4}$, the expression for the phase velocity can be evaluated exactly. It should be noted however,  
that, the same with nonzero scalar mass and/or other interactions present
in the Lagrangian, may lead to a more complex situation and an exact 
analytical result may not be possible. We won't be elaborating on this issue
any further (here), it would be dealt with in a separate publication.

\section{Application:}

The predictions for the class of theories under consideration here, can be tested through optics based experiments set up for laboratory or astrophysical environments. For instance through the observations of the index of the power (of 
the ) spectrum, checking the differential dispersion measure or through the  
measurements of polarization angle, ellipticity at different wavelengths.

\indent
The analysis in this paper, are seemingly more suitable for dispersive and/or 
polarimetric measurements for verifying 
the predictions of these theories.
More over, since we are more interested to find out the astrophysical 
implications of the theory being studied and optics based experiments 
are more suitable for the same, therefore, we will concentrate on 
dispersive or polarimetric analysis here.\\

\subsection{Observables from non-thermal radiation.}
\noindent
In this section we would apply our results for astrophysical situations.
They are of interest because the ambient magnetic field available there,
are many orders of magnitude more, than the same available in laboratory
conditions; also the length of the path the light beam traverses is enormous. 
Keeping this in view we would opt for astrophysical considerations. In astrophysical situations most of the interesting emission mechanisms are of no-thermal 
nature. As the charged particles in these situations accelerate in the ambient 
electromagnetic field, they radiate Electro Magnetic (EM) radiations.

\subsection{Polarized spectrum.}
\noindent
In this section we provide the expression for the mutually orthogonal 
amplitudes of the polarized radiations coming from synchrotron  or 
curvature radiations following [\cite{Ginzburg}- \cite{mena}].

\indent
The amplitudes of the electromagnetic radiation parallel 
or perpendicular to the $\hat{k}$ $\hat{B}$ plane--from the synchrotron or curvature radiation-- are 
given by,
\begin{eqnarray}
A_{\perp} \propto
K_{1/3}\left(\frac{\omega}{2\omega_{sc}}\right) 
\nonumber \\
A_{\parallel}\propto  K_{2/3}\left(\frac{\omega}{2\omega_{sc}}\right) 
\label{power1}
\end{eqnarray}
\noindent
In eqn.(\ref{power1}), $\omega_{sc}=\frac{3}{2}\frac{\Gamma^3}{\rho}$ 
is the peak energy for radiation spectrum with, $\Gamma$, the Lorentz
boost factor for the emitting particles and $\rho$ the radius of curvature
of the particle trajectory. The differential intensity spectrum, given by,
\begin{eqnarray} 
\frac{d^2 I}{d \omega d\Omega}= \frac{(e\omega)^2}{4\pi^2}
\left(|A_{\parallel}|^2+ |A_{\perp}|^2 \right),
\label{intensity}
\end{eqnarray}
 grows as $\omega^{2/3}$  for $\omega \ll \omega_{sc}$,
and falls off exponentially  for $\omega \gg \omega_{sc}$. It is evident
from eqn.(\ref{power1}) that, the EM spectrum from synchrotron or curvature
radiation are emitted in two orthogonally polarized states. \\

\subsection{Dispersive Measures}
In astrophysical situations, dispersive  and polarimetric measures
are very effective  to extract information about a system. For an object
at a distance D we can measure the time taken by the signal to reach us
by measuring $\frac{D}{v_p}$.  That is, by measuring $ t_m= \frac{D}{v_p}$ 
\cite{cfj}.

\indent
In cosmological situations red shift z can be, converted to proper distance 
using (for $\Omega=1$), 
$ D=\frac{2}{3 H_0} \left(1 - \frac{1}{\left(1+z \right)^{3/2}} \right)$, with $H_0$  the Hubble constant being related to to h , by $h=\frac{H_0}
{100 \rm{~Sec}\rm{~Mpc}}km$ and $ h=0.72\pm{0.05} $ obtained from the WMAP data 
\cite{spergel}.
Since we are using natural units, we can take $t=D$.
This in principle can be performed for gamma-ray bursters or pulsar 
observations. In this energy band the  dispersion measure of ( time ) should 
come out to be the proper provided restricts oneself to high energy band, to
avoid the medium induced dispersion effects that's dominant in the low energy
domain.\\

\noindent
In the previous sections we already have mentioned that, for the kind of 
interaction under consideration in this work, the vacuum turns out to be 
birefringent and dichoric for the photons. So the two polarized modes
of photon, propagate with different speed. So, in principle, the orthogonally
polarized signals from an astrophysical object, that originating from the 
source at the same space time point, would reach the observer at two 
different epochs. A similar argument taking into account the effect of
stellar magnetized plasma was initially put forward in \cite{Pdelay1}-\cite{Pdelay4}. However,this argument has been discussed unfavorably in \cite{Pdelay5},
and probably needs more refinement e.g., taking into account the kind of effect
being discussed here and magnetized intergalactic domains, etc. Detailed 
analysis of this effect is beyond the scope  of this paper and would be done 
elsewhere using the techniques of\cite{Jain} .

\subsection{Polarimetric measures}
\noindent
Most of the astrophysical objects are associated with magnetic field, with 
strengths varying from $10^{-9}$ to $10^{13}$ Gauss. Since synchrotron or 
curvature radiation are some of the most efficient non-thermal radiative 
mechanisms, the astrophysical objects mostly radiate non-thermally 
via this processes. A characteristic signature of the radiation coming through
this process is, the radiation is polarized along and perpendicular to the
$\hat{k}$, $\vec{\cal{B}}$ plane; where $\hat{k}$ is the direction vector from 
the source to the observer and $\hat{\cal{B}}$ is the ambient magnetic field 
direction.  The amplitude and the spectrum of the radiation are well known and are discussed later in this paper. 

\indent
The intrinsically polarized nature of the produced radiations turns out to 
be useful to perform polarimetric analysis of the observed data from 
astrophysical sources. The observables for polarimetric analysis are degrees of polarization, linear ($\Pi_{L}$)or circular ($\Pi_c$) and total polarization ($\Pi_{T}$) \cite{born-wolf} . In view of this we would take a digression to the 
essentials of stokes parameters before estimating the polarimetric observables.
\subsubsection{Digression on stokes parameters}

\noindent
In order to evaluate the polarimetric variables (Stokes parameters),
one can construct the coherency matrix by taking different correlations 
of the vector potentials or the fields \cite{born-wolf}. Various optical 
parameters of interest like polarization, ellipticity and  degree of
polarization of a given light beam, can be found from the components of
the coherency matrix constructed from the  correlation functions stated 
above \cite{mybook}. \\

\noindent
For a little digression, the coherency matrix, for  a system with  two degree
of freedom is defined as an ensemble average (where the averaging is done 
over many energy bands)  of  direct product of two vectors:
\begin{eqnarray}
\rho(z)=
\langle \left(
   \begin{matrix}
     \tsi (z) \cr
     \si (z)
   \end{matrix}
\right)
\otimes
\left( {\tsi}(z)\,\,\,  {\si}(z)\right)^{*}
\rangle
=\left(
\begin{matrix}
\langle \tsi(z)\tsi^{*}(z) \rangle  \, \,\,  \langle \tsi(z)\si^{*}(z)\rangle \cr
\langle \tsi^{*}(z)\si(z) \rangle  \, \,\,  \langle \si(z)\si^{*}(z)\rangle
\end{matrix}
\right)
\label{coherencym}
\end{eqnarray}
One important thing to note here, is, under any anticlock-wise rotation by an angle $\alpha$ about an axis i.e., perpendicular
to the vectors $\tsi$ and $\si$, the coherency matrix would transform as:
\begin{eqnarray}
\rho(z)\to \rho^{\prime}(z)=
\langle {\cal{R}}(\alpha)\left(
   \begin{matrix}
     {\tsi}(z) \cr
     {\si}(z)
   \end{matrix}
\right)
\otimes
\left( {\tsi}(z)\,\,\,  {\si}(z)\right)^{*}{\cal{R}}^{-1}(\alpha)
\rangle
\end{eqnarray}
where ${\cal{R}}(\alpha) $ is the rotation matrix. Now from the
relations between the components of the coherency matrix and the stokes parameters:
\begin{eqnarray}
\rm{I}\!\!&=&\!\!<\tsi^{*}(z)\tsi(z)>+<\si^{*}(z)\si(z)>, \nonumber \\
\rm{Q}\!\!&=&\!\!<\tsi^{*}(z)\tsi(z)>- <\si^{*}(z)\si(z)>, \nonumber \\
\rm{U}\!\!&=&\!\!2 \rm{Re}<\tsi^*(z)\si(z)>, \nonumber \\
\rm{V}\!\!&=&\!\!2\,\rm{Im}<\tsi^*(z)\si(z)>.
\label{stokesfield}
\end{eqnarray}\\
It is easy to establish that,
\begin{eqnarray}
\rho(z)= \frac{1}{2} \left(
                 \begin{matrix}
                 {\rm{I}}(z)+~ {\rm{Q}}(z) \,\,\,\,\,\,\,\, {\rm{U}}(z)- i{\rm{V}}(z) \cr
                 {\rm{U}}(z)+i{\rm{V}}(z) \,\,\,\,\,\, {\rm{I}}(z)-~{\rm{Q}}(z)
                 \end{matrix}
\right)
\end{eqnarray}
Therefore, under an anticlock wise rotation by an angle $\alpha$, about an axis perpendicular to the
plane containing $\tsi (z)$ and $ \si(z)$, the density matrix transforms as:
$ \rho(z)~\to~\rho'(z)$; hence the coherency matrix in the rotated frame would be given by,
\begin{eqnarray}
\rho^{\prime}(z)&=&\frac{1}{2}{\cal{R}} (\alpha)  \left(
                 \begin{matrix}
                 {\rm{I}}(z)+~ {\rm{Q}}(z) \,\,\,\,\,\,\,\, {\rm{U}}(z)- i{\rm{V}}(z) \cr
                 {\rm{U}}(z)+i{\rm{V}}(z) \,\,\,\,\,\, {\rm{I}}(z)-~{\rm{Q}}(z)
                 \end{matrix}
\right)
                {\cal{R}}^{-1} (\alpha)~.
\end{eqnarray}
Since for a rotation by an angle $\alpha$--in the anticlock wise direction ( about the axis
that is perpendicular to the plane having $ \tsi $ and $ \si $ on it ) the rotation matrix ${\cal{R}} (\alpha) $
is given by,
\begin{eqnarray}
&{\cal{R}} (\alpha)                =
   \left(
                 \begin{matrix}
                   {\rm{cos}}~{\alpha} \,\,\,\,\,\,\,\,{\rm{sin}}~{\alpha}                                 \cr
                   -{\rm{sin}}~{\alpha} \,\,\,\,\,\,\,\,{\rm{cos}}~{\alpha}
                 \end{matrix}
  \right),
\end{eqnarray}
as a consequence the two stokes parameters $ Q^{\prime}$(z) and $U^{\prime}$ (z), in the rotated frame of reference, would  get related to the same in the unrotated frame, by the relation.
\begin{eqnarray}
\left(
\begin{matrix}
{\rm{Q}}^{'}(z) \cr
{\rm{U}}^{'}(z)
\end{matrix}
\right)
=
\left(
\begin{matrix}
 {\rm{cos}}~{2\alpha} \,\,\,\,\,\,\,\,{\rm{sin}}~{2\alpha}                                 \cr
-{\rm{sin}}~{2\alpha} \,\,\,\,\,\,\,\,{\rm{cos}}~{2\alpha}
\end{matrix}
\right)
\left(
\begin{matrix}
{\rm{Q}}(z) \cr
{\rm{U}}(z)
\end{matrix}
\right)
\end{eqnarray}
\noindent
The other two parameters, i.e., I and V remain unaltered. It is for this reason that
some times I and V are termed invariants under rotation.\\

\noindent
We would like to point out here that, in any frame, the Stokes
parameters are expressed in terms of two angular variables $\chi$ and $\Psi$ usually called
the ellipticity parameter and polarization angle, defined as,
\begin{eqnarray}
{\rm{I}}&=&{\rm{I}}_p \nonumber \\
{\rm{Q}}&=&{\rm{I}}_p {\rm{cos}}~2\Psi ~{\rm{cos}}~2\chi \nonumber \\
{\rm{U}}&=&{\rm{I}}_p {\rm{sin}}~2\Psi ~{\rm{cos}}~2\chi \nonumber \\
{\rm{V}}&=&{\rm{I}}_p  {\rm{sin}}~2\chi.
\label{stokes2}
\end{eqnarray}
The ellipticity angle, $\chi$, following [\ref{stokes2}], can be shown to be equal to,
\begin{eqnarray}
{\rm{tan}}2\chi=\frac{{\rm{V}}}{\sqrt{{\rm{Q}^2}+{\rm{U}}^2}}~,
\end{eqnarray}
\noindent
and the polarization angle can be shown to be equal to.
\begin{eqnarray}
{\rm{tan}}2 \Psi = \frac{\rm{U}}{\rm{Q}}
\label{polangle}
\end{eqnarray}
From the relations given above, it is easy to see that, under the frame rotation,
\begin{eqnarray}
{\cal{R}}(\alpha)
= \left(
\begin{matrix}
 {\rm{cos}}~{2\alpha} \,\,\,\,\,\,\,\,{\rm{sin}}~{2\alpha}                                 \cr
-{\rm{sin}}~{2\alpha} \,\,\,\,\,\,\,\,{\rm{cos}}~{2\alpha}
\end{matrix}
\right)
\end{eqnarray}
the Tangent of $\chi$, i.e., ${\rm{tan}}\chi$ remains invariant, however the tangent of the
polarization angle gets additional increment by twice the rotation angle, i.e.,
\begin{eqnarray}
{\rm{tan}}(2\chi) &&\to {\rm{tan}} (2\chi)  \nonumber \\
{\rm{tan}}(2\Psi)  &&\to {\rm{tan}}(2\alpha+2\Psi).
\end{eqnarray}
\noindent
It is worth noting that the two angles are not quite independent of each other, in fact they are related
to each other. Finally  we end the discussion of use of stokes parameters by noting that, the degree of polarization is usually expressed by,
\begin{eqnarray}
p=\frac{\sqrt{{\rm{Q}}^2+{\rm{U}}^2+{\rm{V}}^2}}{{\rm{I}}_{P_T}}
\end{eqnarray}
where ${\rm{I}}_{P_{T}}$ is the total intensity of the light beam.\\

\noindent
Since we already have the expressions for the stokes parameters in terms of the solutions of the field equations 
(\ref{solnforfields}) 
one can substitute the solutions of the field equations in (\ref{stokesfield}) to arrive at the expressions for $I$, $Q$, $U$ and $V$. The expressions for the same are given by,

\begin{eqnarray}
I(\omega,z)&=&\frac{1}{2} \left[3+ cos\left[\left(\omega_{-}-\omega_{+} \right)z\right]  \right] \\
Q(\omega,z)&=&\frac{1}{2} \left[cos\left[\left(\omega_{-}-\omega_{+} \right)z\right] -1   \right]\\
U(\omega,z)&=& 2.0\left[cos \left[\left(\frac{\omega_{+}-\omega_{-}}{2} \right)z\right]\right] 
\left[cos \left[\left(\frac{\omega_{+}-\omega_{-}}{2} - \omega \right)z\right]\right] \\
V(\omega,z)&=& -2\left[cos \left[\left(\frac{\omega_{+}-\omega_{-}}{2} \right)z\right]\right] 
\left[sin \left[\left(\frac{\omega_{+}-\omega_{-}}{2} + \omega \right)z\right]\right] 
\end{eqnarray}

\section{Astrophysical Accelerators}
\noindent
Standard astrophysical accelerators of charged particles in our 
universe are, white dwarf (WD) pulsar, neutron star (NS) pulsars,
supernova remnants (SNRs), micro-quasars and possibly  gamma-ray 
bursters [\cite{whitedwarf}-\cite{accl22}] to name a few. As the 
charged particles accelerate along the dipole magnetic field lines
\cite{rad} of these astrophysical accelerators, they emit energy through
synchro-curvature radiation [\cite{sc1} -\cite{sc3}], whose 
spectrum (in familiar notations),  peaks at $\omega_{sc}=
\frac{3 \Gamma^3}{2 \rho}$ . 

\indent
For our analysis we would dealing with the emission spectra and polarization studies of compact stars e.g., white dwarf pulsar or neutron star pulsars. We will be following the analysis of  [\cite{sc1} -\cite{sc3}] and 
[\cite{usov1} \cite{usov2}] in this paper. According to the rotating dipole 
model \cite{rad} of compact stars, a strong electric field of magnitude, 
$E_{\parallel}= \frac{\vec{E}.\vec{\cal{B}}}{|\vec{{\cal{B}}}|}$ is generated along the 
magnetic field in the co-rotating magnetosphere of the pulsar. This can accelerate 
particles to ultra high energies \cite{shabad-usov}. During their accelerating 
phase the charged particles emit EM radiation through synchro-curvature 
radiation to be detected by a faraway observer [\!\!\cite{rad}--\cite{fawley}]. 

\indent
A relativistic charged particle of mass m, if moves through a distance $ds$, in 
this electric field, then the change in its energy ( in natural units ), can be written as:

\begin{eqnarray}  
d\left( \Gamma m \right) = \vec{E}_{\parallel}. \vec{ds}= 
- \vec{\nabla} \left(eV\right). \vec{ds}.
\end{eqnarray}
where $V$ is the scalar potential. The solution of the same in one dimension
is provides the value of $\Gamma$ (Lorentz Boost) as a function of position.
It comes out to be: 
\begin{eqnarray}  
 \Gamma (s)  = -\frac{eV (s)}{m} + \rm{constant},
\end{eqnarray}
 where the distance s is measured from the center of the star. The value of the constant is fixed by assuming that on the surface of 
the compact object, velocity of the charged particle was zero, and at the 
surface of the star parallel component of the electric field vanishes,  hence, 
one arrives finally at:
\begin{eqnarray}  
 \Gamma (s)  =1 - \frac{eV(s)}{m}.
\label{gamma-ref}
\end{eqnarray}

\noindent
Equation (\ref{gamma-ref}), is the same as reported in \cite{fawley}. The unity in eqn. (\ref{gamma-ref}) is usually neglected for large Lorentz boost 
\cite{usov1}, however for consistency it has to be retained. Using the relation $V(r)= -\int^r_{0} E_{\parallel}.ds$( when $R$ is the radius
of the star), one can find out the value of $\Gamma(r)$ for a particular position $r$. According to polar cap model, a compact star with surface magnetic field ${\cal{B}}_{s}$, angular velocity $\Omega= \frac{2 \pi}{P}$ (when P is the period), 
would have electric field $ E_{\parallel}$, close to the star surface, given by 
$  E_{\parallel} \simeq 
\frac{{\cal{B}}_s}{192} \left(\frac{\Omega R}{c}\right)^{\frac{5}{2}}\frac{s}
{\Delta R_p} $ for $0\le s \le \Delta R_{p}$ [\cite{usov1}, \cite{arons}]. 
The Polar cap radius is denoted by, $\Delta R_p \simeq \sqrt{\left[\frac{\Omega R}{c}\right]}R$, with $c$ as the velocity of light and is equal to unity according to our system of units.  This relation makes it is easy to observe that, the 
value of the Lorentz boost at a hight $ r= \alpha \Delta R_{p} 
(0\le \alpha \le 1.) $ from the surface of the star is:
\begin{eqnarray}
\Gamma (r)= \frac{1}{16\sqrt{3}}\left(\Omega R \right)^3 \left(\frac{e{\cal{B}}}{m} \right)
R \alpha^2
\end{eqnarray}
when  the distance $r = \alpha \left( \Delta R_P \right)$, i.e., a fraction of 
the polar cap radius from the surface of the star. One can combine this result
with the expression for $\omega_{sc}$, to get a relation between  emission 
energy vs hight. 

\begin{figure}[h!]
\begin{center}
\begin{minipage}[b]{0.47\linewidth}
\centering
\vspace{-2cm}
\includegraphics[width=\textwidth,height=12cm,scale=5.0]{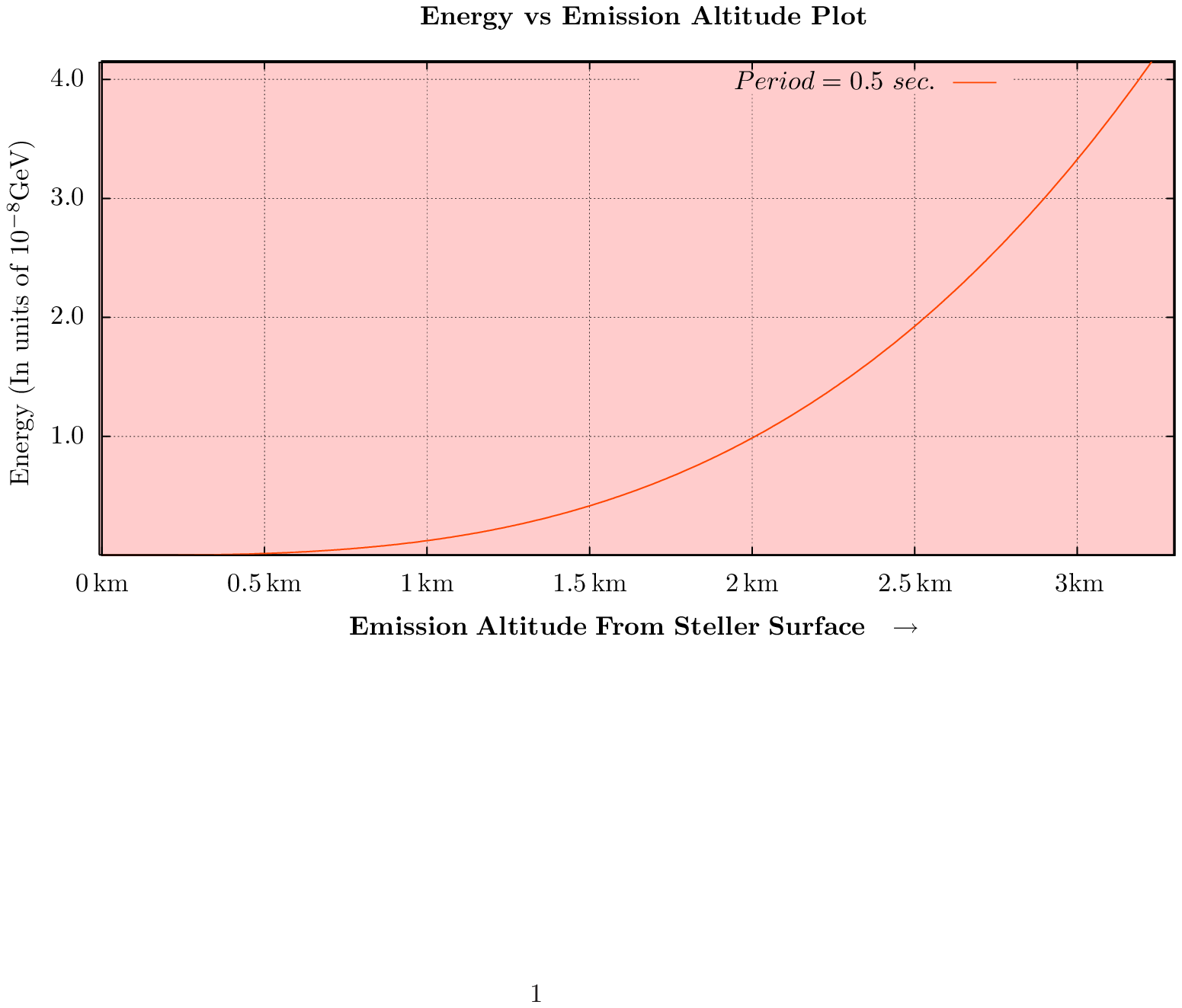}
\vspace{-6cm}
\caption{\small{Plot Of Emission altitude versus energy of emitted synchrotron radiation, for a relatively cold compact star (white dwarf) with dipolar magnetic field strength $\sim 10 ^{13}$, Gauss and period 0.5 sec. }}
\label{figure3}
\end{minipage}
\hspace{0.4cm}
\begin{minipage}[b]{0.47\linewidth}
\centering
\vspace{-2cm}
\includegraphics[width=\textwidth,height=12cm,scale=5.0]{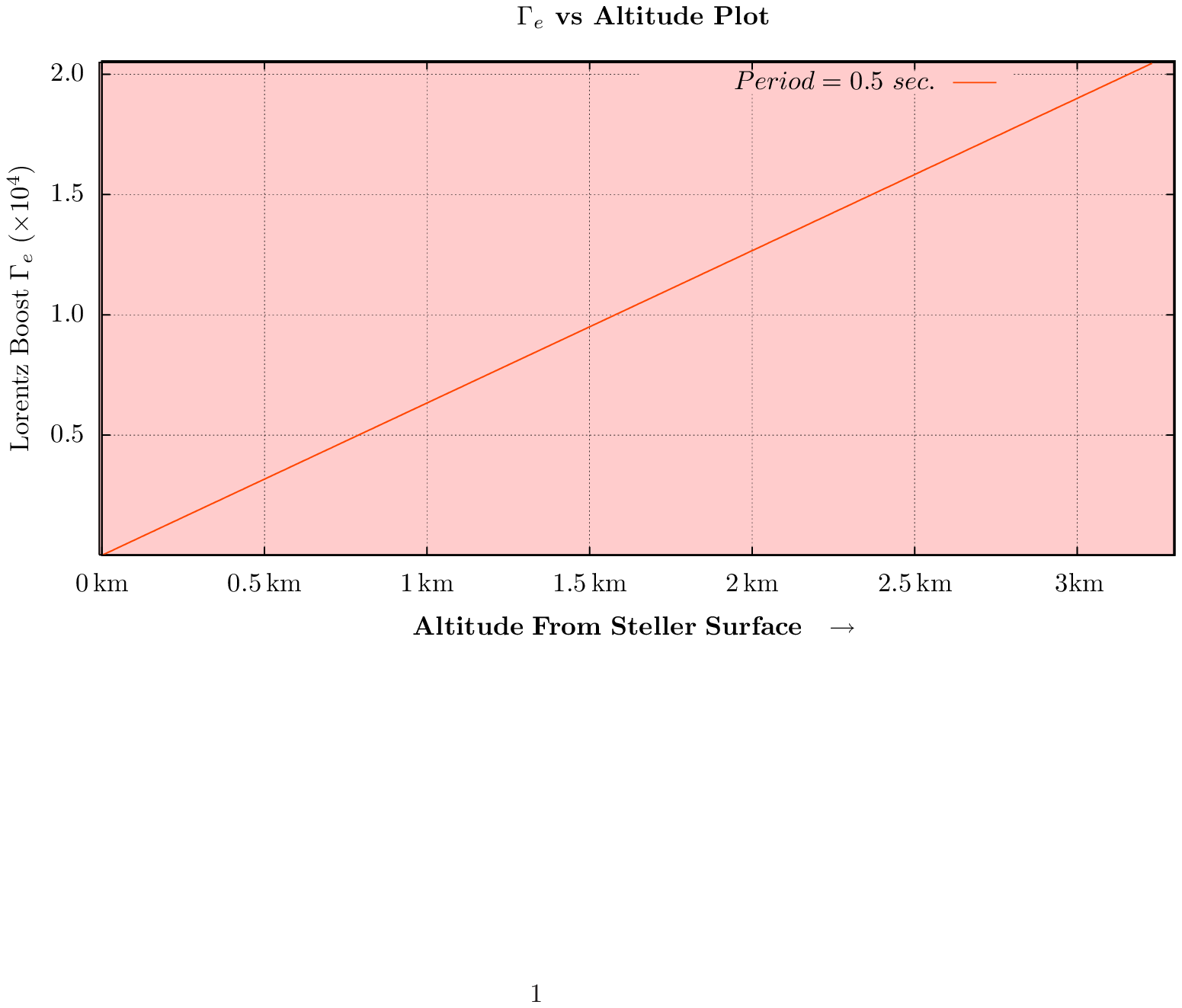}
\vspace{-6cm}
\caption{\small{Plot of Lorentz boost of charged
fermions emitting synchrotron radiation versus emission altitude, for a relatively cold compact star, with same field strength and spin period as in Figure 3.}}
\label{figure4}
\end{minipage}
\end{center}
\end{figure}

\indent
Normally astrophysical compact stars, White Dwarfs (WD) or Neutron Stars (NS ) appear with surface magnetic field strength varying between $10^9$ -$10^{13}$ Gauss. The surface dipolar 
magnetic field strength for young pulsars are more ($ < 10^{9}$ Gauss) than 
the older ones($ > 10^{9}$ Gauss). Assuming the surface dipolar magnetic field 
strength to be around critical magnetic field, i.e $4.4\times 10^{13}$ Gauss 
and a period of 0.5 sec, we have plotted the distance vs boost as well as 
synchrotron emission energy $\omega_{sc}$ in Fig.~[\ref{figure3}] and 
Fig.~[\ref{figure4}] respectively. As can be seen form the plots, that, for a 
compact star of radius 10 km or $10^6$cm, the synchro-curvature radiation reaches the value of $10^{-8}$GeV, 
within 2 km hight from the surface of the star. The same keeps on increasing 
as one moves further away from the surface of the star.\\

\indent
There are two relevant points worth mentioning here, (i) with increase in
the time period $P$ of the compact object, energy of the emitted radiation
at a fixed altitude would tend to increase (ii) we would be assuming that
the photon emission process, takes place close to the last open field 
line and it's quasi tangential to the surface of the star. Although the 
magnetic field strength is expected to vary as 
${\cal{B}}(r)= \frac{{\cal{B}}_S(R) R^2}{r^2}$ 
(a condition that follows from the flux conservation), but because of the 
special emission geometry we have assumed,{\footnote{ Note that this variation 
of ${\cal{B}}$ over photon wave length is not so significant.}} the observables, like $\Psi, \chi$ etc.,  
would not undergo significant variation because of the variation of the
ambient magnetic field.\\
 
\section{Result and Analysis}

\noindent
Earlier we have shown that, the curvature radiation amplitudes for the
plane polarized photons in the magnetized stellar environment follows 
from eqn. [\ref{power1}]. Since our objective in this work to bring
out salient features of such emission, therefore we have assumed the  
initial amplitudes of the two orthogonal polarized modes to be of 
same magnitude. Though this is a simplified assumption, to be true for 
modeling of realistic emission processes taking place in compact astrophysical
objects (WD or NS), however, as it will be clear below, that this is sufficient 
to bring out the new physics issues, those we wish to focus on. \\

\begin{figure}[h!]
\begin{center}
\begin{minipage}[b]{0.45\linewidth}
\centering
\includegraphics[width=\textwidth]{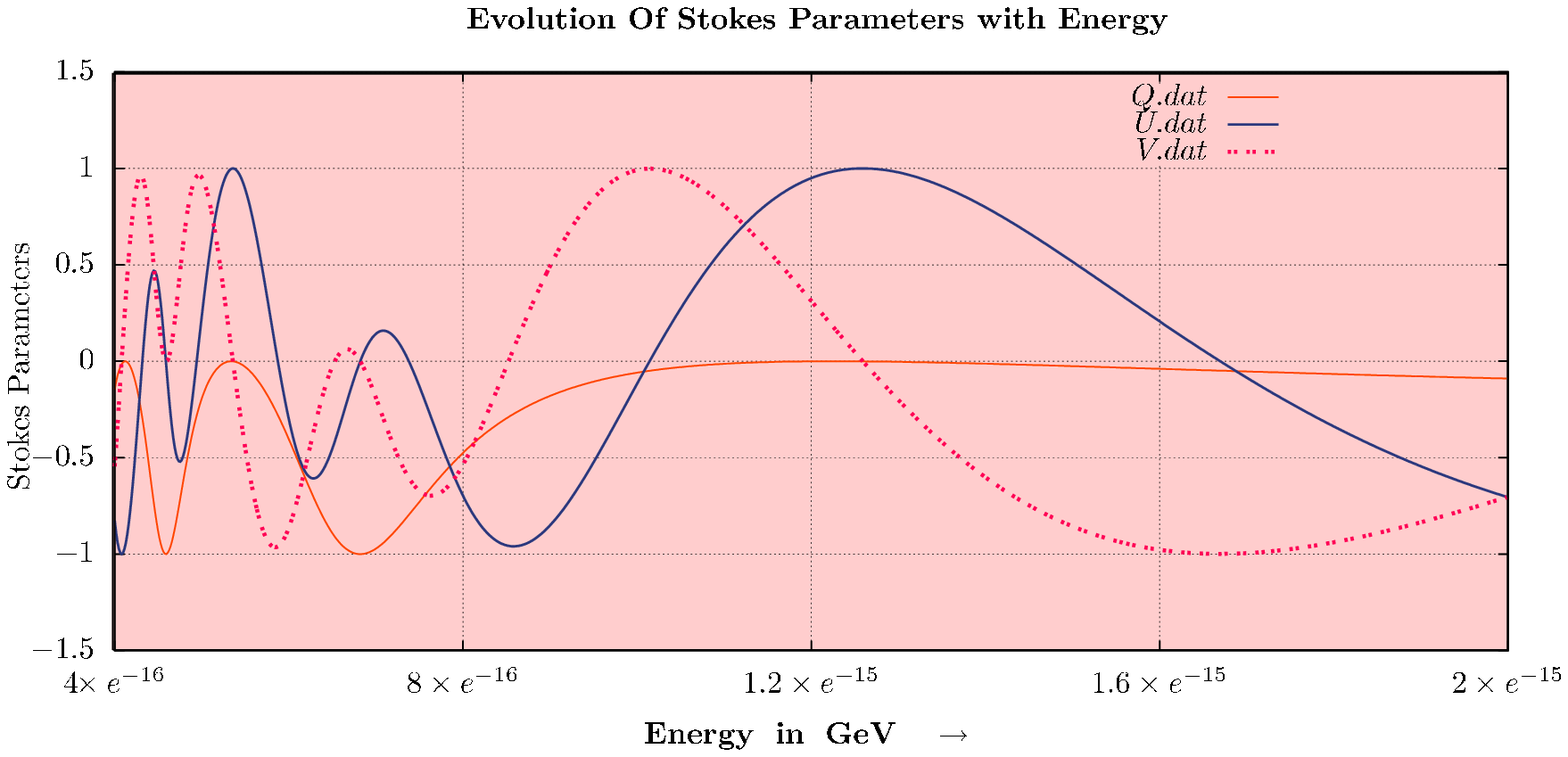}
\caption{Plots for Stokes parameters, Q (orange line), U (blue line) 
and V (dotted orange line), for coupling constant 
$g_{\gamma\gamma\phi}\!=\!10^{-11} GeV^{-1}$
and ${\cal{B}}=4.41 \times 10^{13}$ Gauss--vs $\omega$.}
\label{figure11}
\vspace{-0.5cm}
\end{minipage}
\hspace{0.4cm}
\begin{minipage}[b]{0.45\linewidth}
\centering
\includegraphics[width=\textwidth]{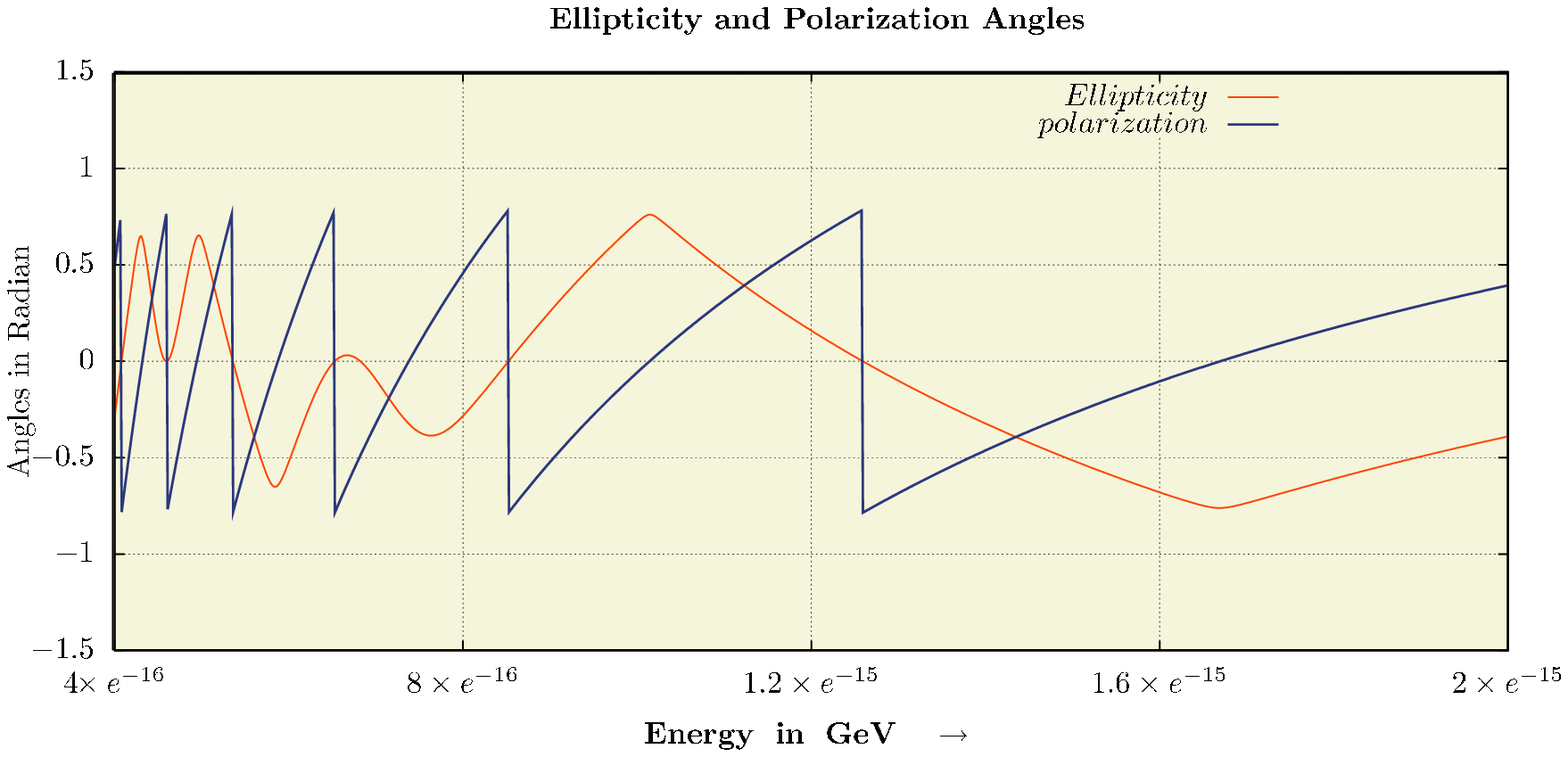}
\caption{Plots for polarization angle $\Psi$ (blue line) and ellipticity 
$\chi$ (orange line), for coupling constant $g_{\gamma\gamma\phi}\!=\!10^{-11} GeV^{-1} $
and ${\cal{B}}=4.41\times 10^{13}$ Gauss--vs $\omega$. }
\label{figure22}
\vspace{-0.445cm}
\end{minipage}
\end{center}
\end{figure}

\noindent
The physics of the optical activity, in this scenario, is the following, as 
the produced electromagnetic beam propagates in the magnetized stellar
environment, the photons with polarization orthogonal to 
the magnetic field, keep mixing with the scalars resulting in a change of 
phase for the same; and the photons with polarization along
the magnetic field propagates freely. The superposition of the two 
decides the net polarization of the system. Since the change of phase is 
dependent on (a) the path traversed by the radiation beam, (b) the strength 
of the ambient magnetic field and (c) the frequency of the photons-- the final 
magnitude of the net polarization of the radiation beam depends 
on all the three. 
 
Since we are interested in the wave propagation in an ambient ( magnetic ) field 
of strength $\sim 10^{13}$ Gauss, believed to exist close to surface of the star, 
we need know the critical synchro-curvature energy $\omega_{sc}$, of the 
emitted photons there in.\\
\indent
The same has been obtained, as already mentioned, using the 
relations of section IV. The critical energy of the emitted photons, as a 
function of altitude from a ultra cold, (WD or NS) pulsar ( with period 
P=0.5 sec.) was evaluated and plotted in Fig.~[\ref{figure3}]. The numerical 
data showed that, at altitudes very close to the surface of the star ( 
$ \le 0.1$ km), the critical energy, turns out to be of the order of 
$O(10^{-16})$ GeV, and it reaches the value of $O(10^{-8})$ GeV, at an 
altitude of $3.5$ km or so.\\
\indent
Therefore assuming the initial amplitudes for both the polarized modes to 
be the same, we have estimated the stokes parameters numerically for 
a frequency range of $4.0 \times 10^{-16}$ GeV to $2.0 \times 10^{-15}$ GeV,
for  $g_{\gamma \gamma c}=\left( 10^{11} GeV\right)^{-1}$ (from PVLAS data 
\cite{pvlas} ) and magnetic field strength ${\cal{B}}=4.41 
\times 10^{13}$ Gauss. The propagation length is taken to be about 
$5\times 10^{5}$ cm ( about $\frac{R_s}{10}$, when $R_s=10^{6}$cm, is the 
stellar radius which is close to that of the astrophysical compact objects). 
The result is plotted in the  Fig.~[\ref{figure11}].\\

\noindent
As can be verified from the initial conditions that, at $z=0$ the stokes 
parameter U is nonzero but Q and V are both zero ( i.e we have linearly 
polarized light ). However the variations of the same, after 
propagation through, a distance $\frac{R_s}{10}$, as a function of energy 
$\omega$ can be seen from figure \label{figure11}. As is clear from the plots,
that at low energy, elliptic polarization, defined by Stokes parameter V, 
though is small in magnitude but the same undergoes modulation with 
increasing $\omega$. Similar behavior is also observed to be taking place
with U. However, the Stokes parameter Q doesn't undergo similar modulation
in magnitude at high energies.\\ 
\indent
It can be checked from the plots that there are situations 
when both Q and V are simultaneously zero except U, 
signaling linear polarization and vice-verse. As the frequency changes, the 
degree of linear polarization decreases and that of circular
polarization increases. This is due to $\gamma$, $\phi$ mixing effect.
We would like to emphasize here that, degree of linear and circular
polarization due to $\phi F_{\mu\nu}F^{\mu\nu}$ coupling need not be of 
very close order at all energies.\\

\indent
The other important observation that we had mentioned already is, that the 
ellipticity $\chi$ and polarization angles $\Psi$ are generally 
multi-valued functions of the energy $\omega$. There can be various values
of  energies $\omega$ at  which the $\Psi$ and $\chi$ could turn out to be 
the same, but mostly they are not, as can be seen from Fig. \ref{figure22}.\\ 
\indent
It should however be noted that, the effects we have discussed so far are 
purely due to mixing, where the polarized multi wavelength beams
are  supposed to have travelled the same distance. This may be achievable 
in laboratory conditions, however,the same may not hold for all class of 
astrophysical objects and all kinds of emissions mechanisms.\\  

\indent
Compact astrophysical objects (stars) usually radiates through 
synchro-curvature radiation, in many energy bands. Energy of an 
emitted photon depends on the emission altitude--measured from 
the surface of the star. For compact stars, this relationship,
is referred at times as altitude energy mapping. The nature of
this mapping depends on the details of the model of the compact 
object. In our illustrative model the low energy photons
originate close to and higher energy photons originate 
far away from the surface of the compact star. Therefore the low 
energy modes would pass through larger distance in the magnetized 
media than the high energy ones.\\ 
\indent
So the kind of polarizations two light beams ( say at two different 
energies $\omega_1$ and $\omega_2$,  with $\omega_1 < \omega_2$), are 
going to acquire would be different once they are out of the stellar 
environment. This is because the two different path lengths traveled 
by the the two beams in the magnetized stellar media, due to emission 
geometry. Since for the kind of physical picture for we have in mind 
( WD, NS or Quasars ), a radiation beam, due to altitude energy effect 
( following dipole emission models ), would be having multiple energies 
where each of these individual components would be traveling different 
path-lengths in the magnetized stellar atmosphere, once out of the 
stellar environment. Therefore, in addition to the $\gamma \phi$ mixing 
induced polarization  effects--an additional path dependent effect would 
also show up at various energy bands of the synchro-curvature radiations 
coming from WD or NS because emission-altitude energy relations, that 
these radiations follow. Hence, for a consistent interpretation of the 
observations, the same needs to be accounted for.\\ 

\subsubsection{Modified Boundary Conditions For Scalars.}
Another important point that we would like to point out here, is,
although the number of scalar particles produced in the 
synchro-curvature emission model of WD or NS are usually zero 
( from kinematical or other considerations ), however, by the time the 
emitted radiation is out of the stellar environment a significant 
amount of scalars may be generated because of the mixing effect. \\
\indent
Therefore, while analyzing synchro-curvature spectra of radiations from 
far away WD or NS, one would need to take this in to consideration for the 
fixing boundary conditions, along with the effect of multiple magnetized
intergalactic domains ( similar model was considered in \cite{TJ}, however
the radiation sources considered there were different).

\section{ Discussion and Outlook.}
In this section we briefly summarize the findings of this work. We have tried 
to point out, in this work, the  important polarimetric signatures in the 
synchro- curvature radiation spectra of cosmologically far away compact 
objects (WD or NS) because of scalar photon mixing.\\

\noindent
Our findings are the following: the mixing effect itself is capable of
producing (a) Elliptic or Circular polarization from initially Plane 
Polarized beams, (b) the amount of or plane, circular or elliptic polarization
generated at different energies need not be of same magnitude at all energy 
bands, for the beams traveling through the same distance and same fields 
strength ( magnetic field strength ${\cal{B}}$), (c) Polarization and Ellipticity angles are
multivalued  functions of energy. There may be several energy bands
where the angles repeat themselves, (d) significant amount of scalars
 may be produced in the beams once they are out of the stellar environment,
(e) the physics of emission of synchro-curvature radiation for these sources, 
makes the monochromatic beams at different  energy bands, travel different 
path lengths in the stellar environment. Hence a phase difference, coming 
from the difference of path travelled by the beams of light at different 
$\omega$, will contribute to their polarimetric signatures. Therefore, 
even with running the risk of repetition, we would like to 
emphasize, that, degree of linear and circular polarization need not be 
close to each other, with dim-5  photon scalar coupling.\\

\noindent
Once the electromagnetic signal is out of the stellar environment, it would 
propagate through the ambient magnetized intersteller, galactic and 
inter-galactic space before reaching the observer. Signals from far 
away objects may also travel through multiple magnetized domains in the 
inter-galactic space. Since EM waves undeergo Faraday Rotation \cite{GKP}
in magnetized enviornment, therefore to find out the contribution of scalar
photon mixing operator to polarimetric data, one needs to estimate further,
the  Faraday rotation induced contribution to the same, following the
procedure discussed in \cite{GJM}. However, this analysis is out of scope 
of current study and  the same will be undertaken else where. \\

\setcounter{equation}{0}
\setcounter{footnote}{0}


\end{document}